\documentclass[twocolumn,prl,showpacs,amsfonts,amsmath,assymb,eufrak]{revtex4-1}
\usepackage{epsfig}
\usepackage{color}
\usepackage{bm}
\usepackage{braket}
\usepackage{graphicx}
\usepackage{amsthm}

\usepackage{calc}

\newcommand{\eq}[1]{\begin{equation} #1 \end{equation}}
\newcommand{\eqa}[2]{\begin{equation} #1 \label{#2} \end{equation}}
\newcommand{\balign}[1]{\begin{align} #1 \end{align}}
\newcommand{\bcases}[1]{\begin{cases} #1 \end{cases}}

\newcommand{\mx}[1]{\begin{pmatrix}#1 \end{pmatrix}}
\newcommand{\dis}{\displaystyle}

\newcommand{\bs}{\boldsymbol}

\newcommand{\figin}[4]
{\begin{figure}[tb]
\centering
\includegraphics[width= #1]{#2.pdf}
\caption{#3}
\label{f:#4}
\end{figure}}

\newcommand{\todayd}{\the\year/\the\month/\the\day}
\newcommand{\del}{\partial}
\newcommand{\bib}{\bibitem}

\newcommand{\mr}{\mathrm}
\newcommand{\lr}{\leftrightarrow}

\newcommand{\lmd}{\lambda}

\newcommand{\lb}{\label}
\newcommand{\nt}{\notag}

\newcommand{\ft}[2]{\left. #1 \right|_{#2}}

\newcommand{\bel}{\begin{easylist}}
\newcommand{\eel}{\end{easylist}}

\newcommand{\eref}[1]{Eq.~\eqref{#1}}

\newcommand{\fref}[1]{Fig.~\ref{f:#1}}

\def \({\left(}
\def \){\right)}
\def \[{\left[}
\def \]{\right]}
\newcommand{\la}{\left\langle}
\newcommand{\ra}{\right\rangle}

\newcommand{\sumtwo}[2]%
{\mathop{\sum_{#1}}_{#2}}
\newcommand{\sumthree}[3]%
{\mathop{\mathop{\sum_{#1}}_{#2}}_{#3}}
\newcommand{\sumfour}[4]%
{\mathop{\mathop{\mathop{\sum_{#1}}_{#2}}_{#3}}_{#4}} 
\newcommand{\prodtwo}[2]%
{\mathop{\prod_{#1}}_{#2}}
\newcommand{\mintwo}[2]%
{\mathop{\min_{#1}}_{#2}}
\newcommand{\maxtwo}[2]%
{\mathop{\max_{#1}}_{#2}}
\newcommand{\maxthree}[3]%
{\mathop{\mathop{\max_{#1}}_{#2}}_{#3}}
\newcommand{\limtwo}[2]%
{\mathop{\lim_{#1}}_{#2}}
\newcommand{\suptwo}[2]%
{\mathop{\sup_{#1}}_{#2}}
\newcommand{\supthree}[3]%
{\mathop{\mathop{\sup_{#1}}_{#2}}_{#3}}
\newcommand{\supfour}[4]%
{\mathop{\mathop{\mathop{\sup_{#1}}_{#2}}_{#3}}_{#4}} 
\newcommand{\inftwo}[2]%
{\mathop{\inf_{#1}}_{#2}}
\newcommand{\infthree}[3]%
{\mathop{\mathop{\inf_{#1}}_{#2}}_{#3}}
\newcommand{\inffour}[4]%
{\mathop{\mathop{\mathop{\inf_{#1}}_{#2}}_{#3}}_{#4}} 

\newcommand\calC{{\cal C}}

\newcommand\calI{{\cal I}}
\newcommand\calJ{{\cal J}}

\newcommand{\bsa}{\boldsymbol{a}}

\newcommand{\bsh}{\boldsymbol{h}}

\newcommand{\bsp}{\boldsymbol{p}}

\newcommand{\bsu}{\boldsymbol{u}}

\newcommand{\bsx}{\boldsymbol{x}}

\newcommand{\Di}{\mathit{\Delta}}

\newcommand{\para}[1]{{\em #1}\/.---}

\def\rnum#1{\resizebox{0.5em}{\height}{\expandafter{\romannumeral #1}}}
\def\Rnum#1{\resizebox{0.5em}{\height}{\uppercase\expandafter{\romannumeral #1}}}

\newcommand{\hsgm}{\hat{\sigma}}

\newcommand{\hX}{\hat{X}}

\newcommand{\htau}{\hat{\tau}}
\newcommand{\dsgm}{\dot{\sigma}}

\newcommand{\pss}{p^{\rm ss}}

\newcommand{\hcalJ}{\hat{\calJ}}
\newcommand{\hcalI}{\hat{\calI}}
\newcommand{\hn}{\hat{n}}

\newcommand{\tlF}{\tilde{F}}

\newcommand{\hcalC}{\hat{\calC}}

\makeatletter
\renewcommand{\@cite}[1]{\textsuperscript{#1)}}
\makeatother

\begin{document}

\preprint{APS/123-QED}

\newcommand{\titlehere}{Time-symmetric current and its fluctuation response relation around nonequilibrium stalling stationary state}

\newcommand{\titlename}{\titlehere}

\preprint{APS/123-QED}

\title{\titlename}

\author{Naoto Shiraishi}
\affiliation{Department of Physics, Gakushuin University, 1-5-1 Mejiro, Toshima-ku, Tokyo 171-8588, Japan}

\date{\today}

\begin{abstract}
We propose a time-symmetric counterpart of the current in stochastic thermodynamics named time-symmetric current.
This quantity is defined with empirical measures and thus is symmetric under time reversal, while its ensemble average reproduces the amount of average current.
We prove that this time-symmetric current satisfies the fluctuation-response relation in the conventional form with sign inversion.
Remarkably, this fluctuation-response relation holds not only around equilibrium states but also around nonequilibrium stationary states if observed currents stall.
The obtained relation also serves as an experimental tool for calculating the value of a bare transition rate by measuring only time-integrated empirical measures.
\end{abstract}

\maketitle


\para{Introduction}
The fluctuation-response relation is one of the most important relations in nonequilibrium statistical mechanics, which reveals a nontrivial connection between the equilibrium fluctuation of currents and the response of current to external perturbations.
This relation was first discovered by Johnson~\cite{Joh28} in experiments of electrical conduction, and then theoretically derived by Nyquist~\cite{Nyq28} from the consistency with the second law of thermodynamics.
The Kubo formula~\cite{Kub57} can be understood as a refinement of the fluctuation-response relation from a microscopic viewpoint.

The extension and generalization of the fluctuation-response relation to nonequilibrium stationary systems have been investigated intensively.
Various formal extensions have been widely investigated by e.g., replacing the time-evolution operator or the initial distribution~\cite{YK67, Aga72, TT74, HT82, PJP09, SS10, CGH21} (see recent reviews~\cite{EMbook, Mar08, CG11, BM13, Mae20} and references therein).
The fluctuation theorem~\cite{ECM93, GC95, Kur98, Mae99, Jar97} is sometimes regarded as a generalization of the fluctuation-response relation to processes far from equilibrium.
Another interesting generalization is seen in the Harada-Sasa relation~\cite{HS05}, in which the violation of the fluctuation-response relation in Langevin systems is directly related to the stationary heat dissipation.
The thermodynamic uncertainty relation~\cite{BS15, Ging16, GRH17} is now understood as a kind of inequality between the fluctuation and the response to perturbations~\cite{DS20, LGU20, KS20, Shi21, DS21}.
The trade-off inequality between heat current and entropy production~\cite{SST16, SS19} can also be understood as a generalization of the fluctuation-response relation to nonequilibrium regimes.
Investigation of equalities and inequalities generalizing the fluctuation-response relation based on stochastic thermodynamics have continued~\cite{APE16, OGH20}.


Previous studies mostly concern the fluctuation and the response of {\it time-antisymmetric quantities} such as currents, which are easily connected to thermodynamic irreversibility.
Instead, we here consider the fluctuation and the response of {\it time-symmetric quantities}.
Two prominent examples of time-symmetric quantities are the empirical measure (staying time at a given state) and activity (the number of jumps), which have been mentioned as key quantities to understand nonequilibrium stationary systems~\cite{RM07, Maes1, Maes2, Mae20}.
Equilibrium systems are characterized solely by the ratio of forward and backward transition rates due to the detailed-balance condition.
By contrast, nonequilibrium stationary systems are sensitive to e.g., coefficients of transition rates and escape rates, which are quantified by time-symmetric quantities.
In fact, several relations for nonequilibrium stationary systems accompany the empirical measure~\cite{DV75, Shi13, BFG15, MNW08, MN14} and the activity~\cite{Maes1, Maes2, Gar07, LAW07, BT12, Shi21}.
A few studies~\cite{MW06, Mae20b} tried to find out fluctuation-response relations on time-symmetric quantities, though these quantities are not so simple.
Despite their importance, the physical role of time-symmetric quantities has still not been fully understood.

In this Letter, we propose a novel time-symmetric quantity, a time-symmetric current, which is constructed with empirical measures.
Despite its time-symmetric property, the time-symmetric current also behaves as a kind of current in that its ensemble average is equal to the conventional current.
We prove that the time-symmetric current satisfies the fluctuation-response relation in the conventional form with sign inversion.
Interestingly, this fluctuation-response relation holds not only around equilibrium systems but also around nonequilibrium stationary systems if the observed current stalls.

The obtained relation can be transformed into an expression of activity, or a bare single transition rate, only in terms of cumulative empirical measures.
Recent experiments on molecular motors~\cite{KSS14, NT21} find that not the ratio of transition rates but the value of a single transition rate plays an important role to understand the specialty of molecular motors.
Our result may help to evaluate a single transition rate and reveal its roles in experiments.

\para{Setup and main result}
We consider stationary continuous-time Markov jump processes on discrete states in $0\leq t\leq \tau$.
We finally take the long-time limit $\tau\to \infty$.
The time evolution of the probability distribution $\bsp$ follows the master equation
\eq{
\frac{d}{dt}\bsp(t)=R\bsp(t),
}
where $R$ is a transition matrix.
We assume the local detailed-balance condition, which leads to the expression of the entropy production rate $\dsgm(t)$ as
\eq{
\dsgm(t)=\sum_{i,j}R_{ij}p_j(t)\ln \frac{R_{ij}p_j(t)}{R_{ji}p_i(t)}.
}
Let $\la \cdot \ra^{\tau}$ be an ensemble average of an observable in the stationary state in the time interval $0\leq t\leq \tau$.
The bracket without superscript $\la \cdot \ra$ represents its long-time average: $\langle \hX\rangle:=\lim_{\tau \to \infty}\frac{1}{\tau}\langle \hX\rangle^\tau$.

The cumulative current from a state $i$ to another state $j$, defined as $\hcalJ_{ij}:=\hn_{ij}-\hn_{ji}$, is a key time-antisymmetric quantity in stochastic thermodynamics, where $\hn_{ij}$ is the number of jumps from the state $j$ to $i$.
Analogously to this, we introduce a key quantity named cumulative {\it time-symmetric current} $\hcalI_{ij}$, which is a current-like but time-symmetric quantity defined as
\eqa{
\hcalI_{ij}:=R_{ij}\htau_j-R_{ji}\htau_i.
}{def-I}
Here, $\htau_i:=\int_0^\tau dt \delta_{w(t), w_i}$ is the empirical staying time (see \fref{EM}).
The current and the time-symmetric current have the same ensemble average
\eq{
J_{ij}:=\langle \hcalJ_{ij}\rangle=\langle \hcalI_{ij}\rangle=:I_{ij}.
}
On the other hand, under the time-reversal of trajectories ($\Gamma\to \Gamma^\dagger$), the current changes its sign, $\calJ(\Gamma^\dagger)=-\calJ(\Gamma)$, while the time-symmetric current is kept invariant, $\calI(\Gamma^\dagger)=\calI(\Gamma)$.

\figin{8.5cm}{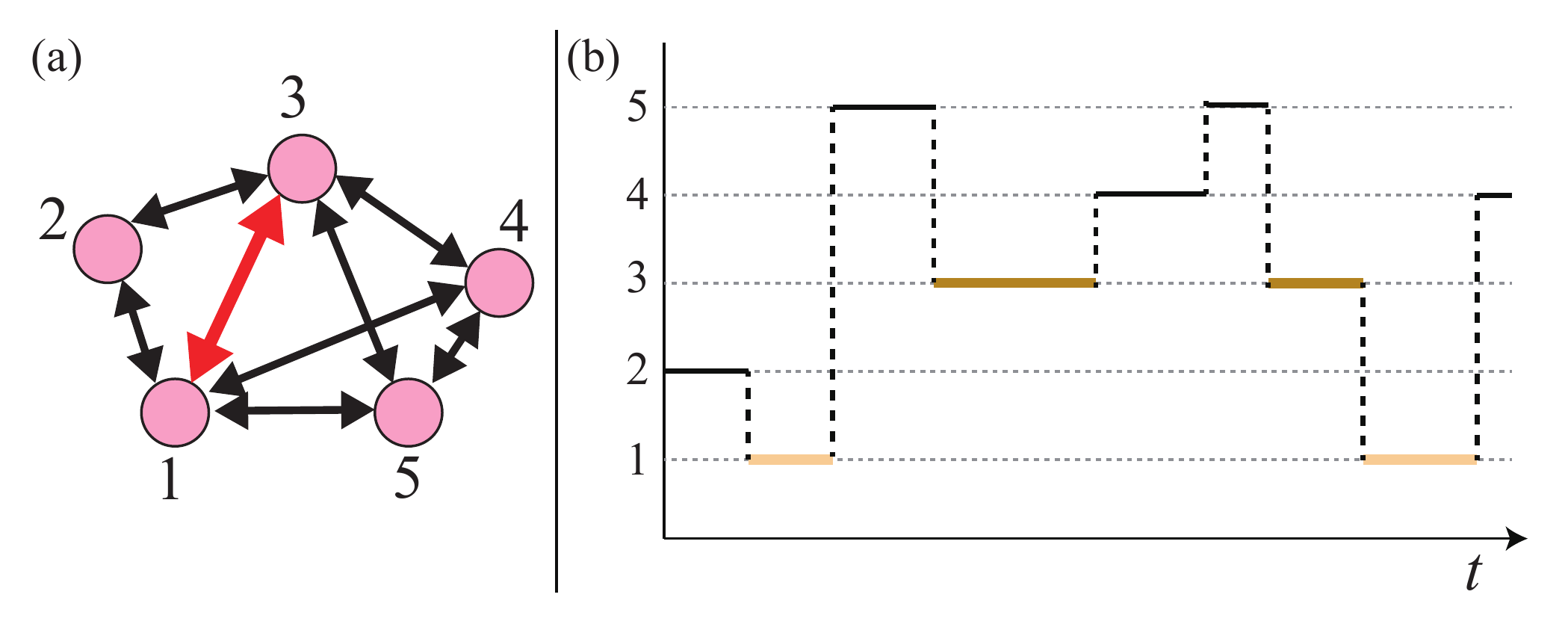}{
(a): Example of a state space of Markov jump processes on five states labeled from 1 to 5.
We here consider $\hcalI_{13}$.
The corresponding edge is drawn in red.
(b): Example of a trajectory $\Gamma$ in this Markov process.
The sums of lines with dark orange and light orange represent $\htau_3$ and $\htau_1$, respectively.
Using these two quantities, the value of $\hcalI_{13}$ is determined.
}{EM}

We consider time-symmetric thermodynamic currents $I_a:=\sum_{(i,j)}d^a_{ij}I_{ij}=\sum_{(i,j)}d^a_{ij}J_{ij}$ with general weights $d^a_{ij}$.
Its conjugate force $h_a$ satisfies $\dsgm=\sum_a h_aI_a=\sum_a h_aJ_a$, where we set the Boltzmann constant to unity and $a$ is a label of the types of currents.
Note that $\hcalI_{ij}$ and $\hcalI_a:=\sum_{(i,j)}d^a_{ij}\hcalI_{ij}$ depend on $\bsh$ through the transition rates $R_{ij}$ and $R_{ji}$.
Manifesting the $\bsh$-dependence of them, we also express $\hcalI_{ij, \bsh}$ and $\hcalI_{a,\bsh}$.

The conventional fluctuation-response relation reads $\del_{h_a} J_a=\frac12 \langle (\hcalJ_a)^2 \rangle$ around equilibrium states, and is recently extended to nonequilibrium stationary states with stalling $J_a$ in a microscopic sense (i.e., $J_{ij}=0$ for all $i,j$ with nonzero $d^a_{ij}$)~\cite{APE16}.
Analogously, we find a fluctuation-response relation for a time-symmetric current
\eqa{
\ft{\frac{\del \langle \hcalI_{a, \bsh^*}\rangle_{\bsh}}{\del h_a}}{\bsh=\bsh^*}=-\frac12 \langle (\hcalI_{a, \bsh^*})^2\rangle_{\bsh^*},
}{h-main}
which is valid around any equilibrium states and nonequilibrium stalling states.
This is our main result of this Letter.
Here, $\bsh^*$ is a parameter where the current $J_a$ stalls in the microscopic sense, and $\langle \cdot \rangle_{\bsh}$ is the stationary ensemble average with the parameter $\bsh$.
We remark that the left derivative only concerns the path probabilities and does not touch the observable $\hcalI_a$.
The relation \eqref{h-main} suggests that the time-symmetric current defined as \eref{def-I} is a reasonable time-symmetric counterpart of the current $J$.

\figin{7.5cm}{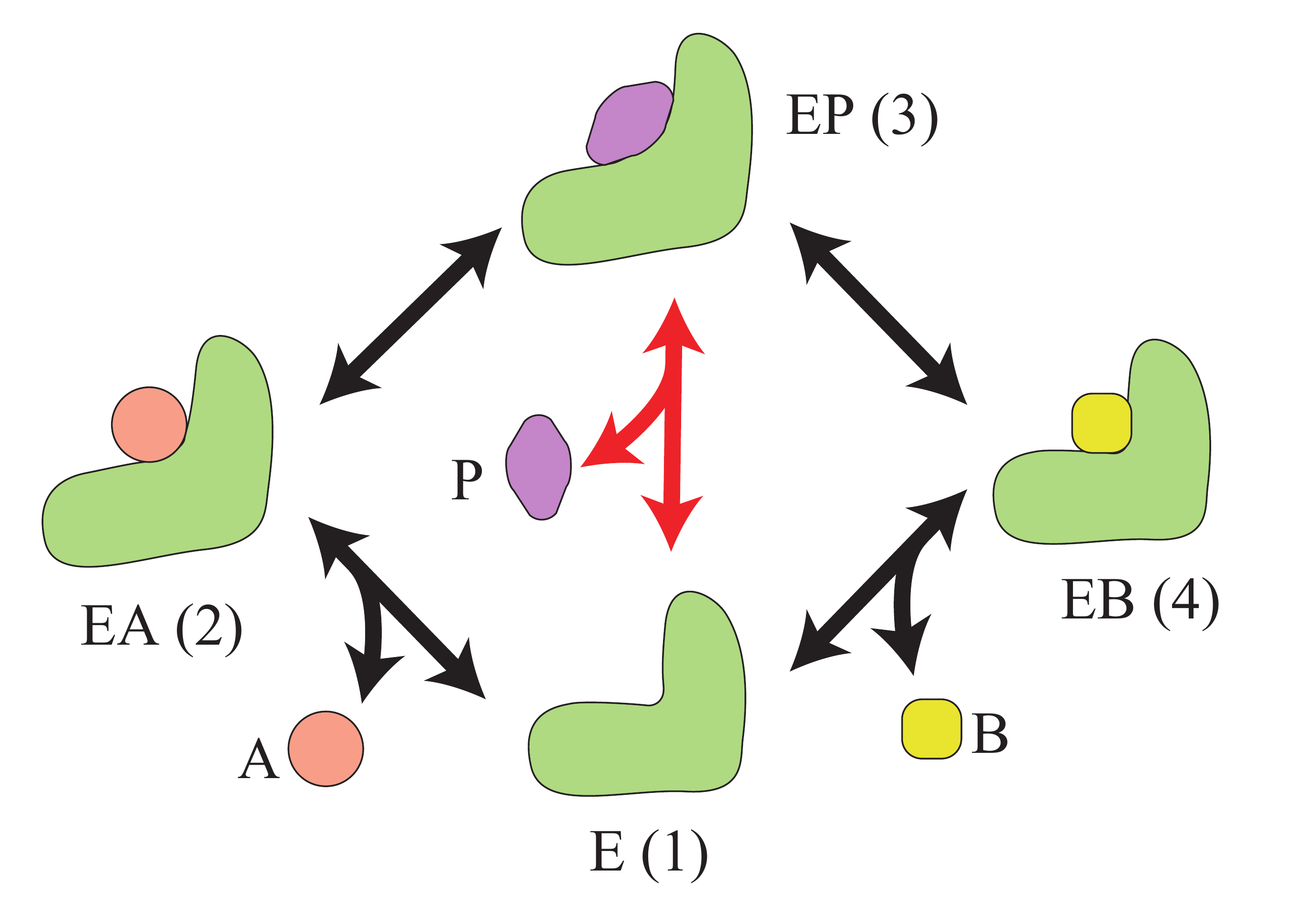}{
State space of the toy model.
An enzyme E involves two different reaction paths: E$\lr$EA$\lr$EP$\lr$E and E$\lr$EB$\lr$EP$\lr$E.
Two substances, A and B, and a product, P, are supplied from three different chemical baths.
We also label the four states, E, EA, EP, and EB as 1,2,3, and 4.
}{toy}

\para{Demonstration with a toy model}
We here numerically demonstrate the validity of Eq.~\eqref{h-main} in a simple stationary chemical-reaction model.
Consider an enzyme E which generates a product P in two different ways: consuming a substrate A or another substrate B in chemical baths.
Two reaction networks are expressed as (see \fref{toy})
\begin{quote}
\centering
E(+A) $\lr$ EA $\lr$ EP $\lr$ E(+P), \\
E(+B) $\lr$ EB $\lr$ EP $\lr$ E(+P).
\end{quote}
The reactions E(+A)$\lr$EA, E(+B)$\lr$EB, and E(+P)$\lr$EP are driven by the chemical baths of A, B, and P, respectively, and the reactions EA$\lr$EP and EB$\lr$EP are driven by a heat bath.
For brevity, we label four states E, EA, EP, and EB as 1, 2, 3, and 4, respectively.

We set transition rates of this model as $R_{21}=2$, $R_{12}=2$, $R_{32}=R_{23}=k$, $R_{31}=1$, $R_{13}=2e^{-h/2}$, $R_{43}=3$, $R_{34}=2$, $R_{41}=r$, and $R_{14}=2$ with $h=0$ and take the time-scale separation limit $k\to \infty$ first, so that the probability distribution between states 2 and 3 is always in local equilibrium.
We examine the relation \eqref{h-main} for a time-symmetric current from 1 to 3.
We compute three quantities; the minus of the left-hand side and the right-hand side of \eref{h-main} (i.e., $\frac12 \langle \hcalI_{31}^2\rangle_{\bsh^*}$ and $-\del_{x_{31}}\langle \hcalI_{31,\bsh}\rangle_{\bsh^*}$), and the current from $1$ to $3$ with changing the value of $r$.

The results are plotted in \fref{plot}, where the orange, blue, and green graphs represent the values of $\frac12 \langle \hcalI_{31}^2\rangle_{\bsh^*}$, $-\del_{x_{31}}\langle \hcalI_{31,\bsh}\rangle_{\bsh^*}$, and $J_{31}$, respectively.
These graphs clearly show that although both sides of \eref{h-main} take different values in general, they coincide if the current between $1$ and $3$ vanishes (i.e., $r=3$; the dashed line in \fref{plot}), which confirms the validity of \eref{h-main}.
We remark that the system is in a nonequilibrium stationary state, not in equilibrium, at $r=3$.

\figin{8.5cm}{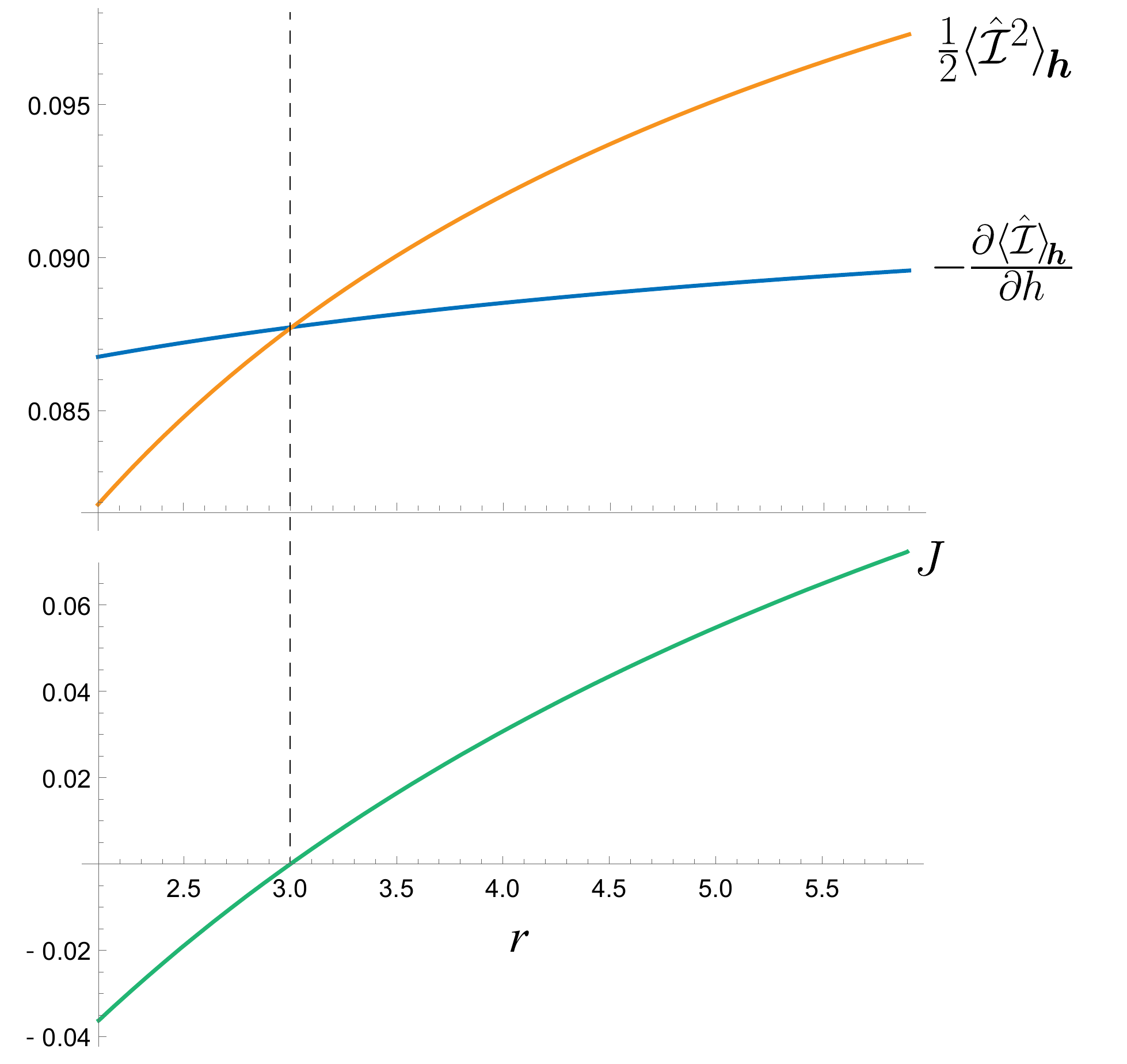}{
Plots of $\frac12 \langle \hcalI_{31}^2\rangle_{\bsh^*}$(orange), $-\del_{x_{31}}\langle \hcalI_{31,\bsh}\rangle_{\bsh^*}$(blue), and $J_{31}$(green) with respect to a parameter $r$ in the toy model.
The edge between 1 and 3 stalls at $k=3$, and at this time $\frac12 \langle \hcalI_{31}^2\rangle_{\bsh^*}$ and $-\del_{x_{31}}\langle \hcalI_{31,\bsh}\rangle_{\bsh^*}$ take the same value.
}{plot}

\para{Theoretical background of \eref{h-main}}
The fluctuation-response relation \eqref{h-main} is derived from that for a single edge and double edges, which are the fundamental relations in the theoretical aspect.
We suppose that we can solely change the strength of driving from $i$ to $j$, that is, the transition rates $R_{ji }$ and $R_{ij}$.
We employ a set of parameters on  each edge $\bsx$ such that $\del _{x_{ij}}\ln ({R_{ij}}/{R_{ji }})=1$.
The time-symmetric current $\hcalI_{ij, \bsx}$ is defined in a similar manner to $\hcalI_{ij, \bsh}$.

Let $\bsx^*$ be a parameter with which the probability current between $i$ and $j$ stalls: $J_{ij}(\bsx^*)=R_{ij}(\bsx^*)\pss_{j}(\bsx^*)-R_{ji }(\bsx^*)\pss_{i}(\bsx^*)=0$.
Here, $\bsp^{\rm ss}(\bsx)$ is the stationary probability distribution at the parameter $\bsx$.
We emphasize that the system at $\bsx=\bsx^*$ may have finite currents except the edge $ij$ and thus is in a highly nonequilibrium steady state in general.
In this setting, we have the following fluctuation-response relation around the stalling state with $\bsx=\bsx^*$:
\balign{
\ft{\frac{\del \langle \hcalI_{ij, \bsx^*}\rangle_{\bsx}}{\del x_{ij}}}{\bsx=\bsx^*}&=-\frac12 \langle \hcalI_{ij, \bsx^*}^2\rangle_{\bsx^*}, \lb{I-main} \\
\frac{\del \langle \hcalI_{kl, \bsx^*}\rangle_{\bsx}}{\del x_{ij}}+\ft{\frac{\del \langle \hcalI_{ij, \bsx^*}\rangle_{\bsx}}{\del x_{kl}}}{\bsx=\bsx^*}&=- \langle \hcalI_{ij, \bsx^*}\hcalI_{kl, \bsx^*}\rangle_{\bsx^*}. \lb{II-main}
}
Using ${\del_{ h_a}}\ln (R_{ij}/R_{ji})=d^a_{ij}$, above relations directly imply the main result \eqref{h-main}.
We shall prove Eqs.~\eqref{I-main} and \eqref{II-main} at the end of this Letter.
Note that if the system is around equilibrium, the counterpart of the Onsager reciprocity relation $\del_{x_{ij}}\langle \hcalI_{kl, \bsx^*}\rangle_{\bsx}=\del_{x_{kl}} \langle \hcalI_{ij, \bsx^*}\rangle_{\bsx}$ is also satisfied.

\para{Determining bare transition rate and activity}
The obtained fluctuation-response relation \eqref{I-main} leads to the detection of a bare transition rate (and activity) by quantifying empirical measures.
To demonstrate this, we construct the following quantity named as {\it twisted empirical measure}
\eq{
\hcalC_{ij, \bsx}:=\frac{\htau_{j}}{\pss_{j}(\bsx)}-\frac{\htau_{i}}{\pss_{i}(\bsx)},
}
which quantifies the difference of empirical measures relative to its average between the states $i$ and $j$.
At stalling states, $R_{ji }(\bsx^*)\pss_{i}(\bsx^*)=R_{ij}(\bsx^*)\pss_{j}(\bsx^*)$ is satisfied, and thus the twisted empirical measure is connected to the time-symmetric current as
\eqa{
\hcalI_{ij,\bsx^*}=R_{ij}(\bsx^*)\pss_{j}(\bsx^*)\hcalC_{ij, \bsx^*}.
}{IC-connect}
Note that $R_{ij}(\bsx^*)\pss_{j}(\bsx^*)$ is the half of the activity, or traffic, $A_{ij}(\bsx^*):=R_{ji }(\bsx^*)\pss_{i}(\bsx^*)+R_{ij}(\bsx^*)\pss_{j}(\bsx^*)$ in case of stalling.

We consider the response of a twisted empirical measure ${\langle \hcalC_{ij, \bsx^*}\rangle'}_{\bsx=\bsx^*}:=\ft{\del_{x_{ij}} \langle \hcalC_{ij, \bsx^*}\rangle_{\bsx}}{\bsx=\bsx^*}$, which is equal to the derivative of the stochastic entropy difference between the states $i$ and $j$:
\eqa{
{\langle \hcalC_{ij, \bsx^*}\rangle'}_{\bsx=\bsx^*}=\ft{\del_{x_{ij}} \ln \(\frac{\pss_{j}(\bsx)}{\pss_{i}(\bsx)}\)}{\bsx=\bsx^*}.
}{dif-stoc-ent}
Then, the fluctuation-response relation \eqref{I-main} is also written as
\eqa{
\frac{{\langle \hcalC_{ij, \bsx^*}\rangle'}_{\bsx=\bsx^*}}{ \langle \hcalC_{ij, \bsx^*}^2\rangle_{\bsx^*}}=\frac12 R_{ij}(\bsx^*)\pss_{j}(\bsx^*)=\frac14 A_{ij}(\bsx^*).
}{main-stall}
By dividing both sides by $\pss_{j}(\bsx^*)$, the left-hand side consists of only empirical measures, and the right-hand side becomes the bare transition rate $R_{ij}(\bsx^*)$.
In other words, this relation provides an expression of a bare transition rate only in terms of empirical measures.
In the experimental aspect, this bridges a bare transition rate $R_{ij}$, which is usually considered to be measured only through real-time observations, to a cumulative quantity with empirical measures $\calC$.
This relation may help experimental detection of transition rates, where they are intractable through direct measurements.
We note that even if systems of our interest are not stalling, as long as we can tune other transition rates with keeping the transition rates of our interest $R_{ij}$ and $R_{ji}$ unchanged, then we can modify other transition rates to make the edge $ij$ stalling and successfully apply \eref{main-stall} to calculate $R_{ij}$.

\para{Proof of Eqs.~\eqref{I-main} and \eqref{II-main}}
To prove Eqs.~\eqref{I-main} and \eqref{II-main}, we employ two different partial entropy productions; the original, or passive, partial entropy production~\cite{SS15} and the informed one~\cite{PE17, Bis17}.
The partial entropy production is a generalization of entropy production to a subset of transitions, which has various applications from information thermodynamics~\cite{SIKS15, SMS16}, the efficiency of autonomous engines ~\cite{Shi15, Shi17}, to the inference of dissipation~\cite{Bis17, Mar19}.
Both partial entropy productions satisfy fluctuation theorems, which lead to two different fluctuation-response relations on currents~\cite{SS15, APE16}.
For self-containedness, we provide the proofs of these two fluctuation theorems in the Supplementary material~\cite{Supple}.
We shall derive Eqs.~\eqref{I-main} and \eqref{II-main} by considering the difference between these two fluctuation-response relations.

We start from the following partial entropy production with setting observed transitions $\Omega$ as the edges $ij$ and $kl$:
\balign{
\hsgm_{\Omega}:=&\sum_{e=(i,j),(k,l)}a_e(\bsx)\hcalJ_e-J_e(\bsx)\hcalC_{e,\bsx},
}
where $a_{ij}(\bsx):=\ln ({R_{ij}(\bsx)\pss_{j}(\bsx)}/{R_{ji }(\bsx)\pss_{i}(\bsx)})$ represents the total force associated with this edge.
This partial entropy production satisfies the fluctuation theorem~\cite{SS15, Supple}
\eqa{
\la e^{-\hsgm_{\Omega}}\ra_{\bsx}^\tau=1
}{SS-FT}
for the stationary system with parameter $\bsx$.

Following the standard technique to derive the fluctuation-response relation from the fluctuation theorem~\cite{Nak11} (see also \cite{Supple}), we expand \eref{SS-FT} with $\bsa$ around the stalling state ($\bsx=\bsx^*$).
Here we note that $a_{ij}(\bsx)$ is a small parameter around the stalling state of the order of $\Di \bsx:=\bsx-\bsx^*$.
With keeping in mind $J_{ij}(\bsx)\hcalC_{ij, \bsx}=(e^{a_{ij}(\bsx)}-1)R_{ji}(\bsx)\pss_i(\bsx)\hcalC_{ij, \bsx}=a_{ij}(\bsx)\hcalI_{ij,\bsx}+O(a_{ij}(\bsx)^2)$, the coefficients of $a_{ij}^2$ and $a_{ij}a_{kl}$ in this expansion yield
\balign{
\ft{\frac{\del J_{ij}(\bsa)}{\del a_{ij}}}{\bsa={\bs 0}}&=\frac12 (\langle \hcalJ_{ij}^2\rangle_{\bsx^*}+\langle \hcalI_{ij, \bsx^*}^2\rangle_{\bsx^*})={R_{ij}(\bsx^*)\pss_{j}(\bsx^*)}, \lb{PEP1} \\
0&=\langle \hcalJ_{ij}\hcalJ_{kl}\rangle_{\bsx^*}+\langle\hcalI_{ij, \bsx^*}\hcalI_{kl, \bsx^*}\rangle_{\bsx^*}, \lb{PEP2}
}
respectively.
We here used two relations:
First is $\del_{a_{kl}}J_{ij}=\del_{a_{ij}}J_{kl}=0$ and $\del_{x_{ij}}J_{ij}=R_{ij}\pss_j$ around the stalling states, which follows from $J_{ij}=a_{ij}R_{ij}\pss_j+O(\{ a_{ij}, a_{kl}\} ^2)$.
Second is the following relation
\eqa{
\langle \hcalI_{ij, \bsx^*}\hcalJ_{kl}\rangle_{\bsx^*}+\langle \hcalI_{kl, \bsx^*}\hcalJ_{ij}\rangle_{\bsx^*}=0,
}{IJ-zero}
which holds including the case with $i=k$ and $j=l$.
This relation can be easily shown in equilibrium states by using the time-reversal symmetry.
In contrast, it is not easy to prove \eref{IJ-zero} in the nonequilibrium stalling states.
We prove it by applying the method of generating functions and the perturbation expansion of eigenvalues, which is presented in the Supplementary material~\cite{Supple}.

Next, the informed partial entropy production with the same $\Omega$
\eq{
\hsgm_\Omega^{\rm I}:=\hcalJ_{ij}\ln \frac{R_{ij}(\bsx)\pss_{j}(\bsx^*)}{R_{ji }(\bsx)\pss_{i}(\bsx^*)}+\hcalJ_{kl}\ln \frac{R_{kl}(\bsx)\pss_{l}(\bsx^*)}{R_{lk}(\bsx)\pss_{k}(\bsx^*)}
}
satisfies the fluctuation theorem~\cite{PE17, Supple}
\eqa{
\langle e^{-\hsgm_\Omega^{\rm I}}\rangle_{\bsx|\bsp(\bsx^*)}^\tau =1,
}{APE-FT}
where $\langle\cdot \rangle_{\bsx|\bsp(\bsx^*)}^\tau$ is the ensemble average with the transition rate $R(\bsx)$ and the initial distribution $\bsp(\bsx^*)$.
Remark that the coefficient of $\hcalJ_{ij}$ is slightly different from $a(\bsx)$ in that the transition rates (i.e., the heat term) is fixed at the stalling state $\bsx^*$, not the actual transition rates with $\bsx$.
We expand it with $\Di \bsx$ around the stalling state with taking the long-time limit $\tau \to \infty$, whose coefficients of $\Di x_{ij}^2$ and $\Di x_{ij}\Di x_{kl}$ yield
\balign{
\ft{\frac{\del J_{ij}(\bsx)}{\del x_{ij}}}{\bsx=\bsx^*}&=\frac12 \langle \hcalJ_{ij}^2\rangle_{\bsx^*}, \lb{APE1} \\
\frac{\del J_{kl}(\bsx)}{\del x_{ij}}+\ft{\frac{\del J_{ij}(\bsx)}{\del x_{kl}}}{\bsx=\bsx^*}&=\langle \hcalJ_{ij}\hcalJ_{kl}\rangle_{\bsx^*}. \lb{APE2}
}

The Jacobian matrix from $\bsx$ to $\bsa$ at the stalling state $\bsx=\bsx^*$ is expressed as
\eqa{
\ft{\frac{\del a_{kl}}{\del x_{ij}}}{\bsx=\bsx^*}=\delta_{(i,j), (k,l)}+\frac{1}{R_{kl}(\bsx^*)\pss_l(\bsx^*)}\ft{\frac{\del \langle \hcalI_{kl, \bsx^*}\rangle_{\bsx}}{\del x_{ij}}}{\bsx=\bsx^*},
}{axJacobian}
which follows from $a_{kl}=x_{kl}+\ln (p_l/p_k)$ and Eqs.~\eqref{IC-connect} and \eqref{dif-stoc-ent}.
Finally, we transform Eqs.~\eqref{APE1} and \eqref{APE2} from the derivative with $\bsx$ to that with $\bsa$ as $\del_{x_{ij}}=(\del_{x_{ij}} a_{ij})\del_{a_{ij}}+(\del_{x_{ij}}a_{kl})\del_{a_{kl}}$ with plugging Eqs~\eqref{axJacobian} and \eqref{PEP1}, and then subtract Eqs.~\eqref{PEP1} and \eqref{PEP2} respectively, which leads to the desired relations \eqref{I-main} and \eqref{II-main}.

\para{Discussion}
We introduced the time-symmetric current, whose ensemble average coincides with the conventional current while it is time-symmetric.
We showed that this symmetric current satisfies the fluctuation-response relation \eqref{h-main} with the same form as the conventional one with sign inversion.
This fluctuation-response relation holds not only around equilibrium states but also around nonequilibrium stationary states if the observed edges stall.
Experimental verifications and applications of this relation will be fruitful tasks.

One big remaining question is the deep physical meaning of the time-symmetric current $\calI$.
We discussed the similarity of the time-symmetric current to the conventional current, while its time-symmetric face will part ways with the conventional current.
With recalling the importance of time-symmetric quantities in nonequilibrium stationary states, our results, together with the investigation of the specialty of time-symmetric properties, will serve as an important step toward the full understanding of nonequilibrium physics.

\bigskip

\para{Acknowledgement}
The author thanks Keiji Saito for fruitful discussion.
The author thanks Takayuki Ariga, Christian Maes, Yohei Nakayama, and Shin-ichi Sasa for helpful comments.
NS is supported by JSPS KAKENHI Grants-in-Aid for Early-Career Scientists Grant Number JP19K14615.



\clearpage

\pagestyle{empty}

\makeatletter
\long\def\@makecaption#1#2{{
\advance\leftskip1cm
\advance\rightskip1cm
\vskip\abovecaptionskip
\sbox\@tempboxa{#1: #2}%
\ifdim \wd\@tempboxa >\hsize
 #1: #2\par
\else
\global \@minipagefalse
\hb@xt@\hsize{\hfil\box\@tempboxa\hfil}%
\fi
\vskip\belowcaptionskip}}
\makeatother
\newcommand{\vo}{\upsilon}
\newcommand{\midskip}{\vspace{3pt}}

\setcounter{equation}{0}
\def\theequation{A.\arabic{equation}}

\begin{widetext}

\begin{center}
{\large \bf Supplemental Material for  \protect \\ 
  ``\titlehere'' }\\
\vspace*{0.3cm}
Naoto Shiraishi$^{1}$  \\
\vspace*{0.1cm}

{$^{1}$Department of Physics, Gakushuin University, 1-5-1 Mejiro, Toshima-ku, Tokyo 171-8588, Japan} 

\end{center}

\setcounter{equation}{0}
\renewcommand{\theequation}{S.\arabic{equation}}

In this Supplemental Material, we provide proof of a relation whose proof is not presented in the main text.
In addition, for the self-containedness of this manuscript, we provide the derivation of two fluctuation theorems for partial entropy productions, whose proofs are presented in Ref.~\cite{SS15} and Refs.~\cite{APE16, PE17}, and the derivation of fluctuation-response relation from the fluctuation theorem.
This Supplemental Material has reference numbers in common with the main text.


\bigskip\noindent
{\bf \large A. Proof of \eref{IJ-zero}}
\midskip

Here, we prove \eref{IJ-zero} ($\langle \hcalJ_{ij} \hcalI_{kl, \bsx^*}\rangle_{\bsx^*}+\langle \hcalI_{kl, \bsx^*}\hcalJ_{ij}\rangle_{\bsx^*}=0$) under the stalling condition $J_{ij}(\bsx^*)=J_{kl}(\bsx^*)=0$.
More precisely, we consider a discrete-time Markov chain given by $\bsp(n+1)=T\bsp(n)$ with a diagonalizable transition matrix $T$, and prove $\langle \hcalJ_{ij} \hcalI_{kl, \bsx^*}\rangle_{\bsx^*}+\langle \hcalI_{kl, \bsx^*}\hcalJ_{ij}\rangle_{\bsx^*}=0$ for this process.
To reduce the above result to continuous-time Markov jump processes, we set $T_{ij}=\delta_{ij}+R_{ij}\Di t$ and take $\Di t\to 0$ limit.
Since diagonalizable matrices are dense in the space of matrices, our claim is shown to be valid for any transition matrices $T$ by resorting to the continuity of quantities with respect to the change in matrix elements.

We treat the case of $i=k$ and $j=l$ in detail, and then explain the case of $i=k$ but $j\neq l$.
Since the case of $i\neq k$ and $j\neq l$ is essentially the same as the above latter case, we just comment on this case briefly.

\bigskip

\underline{\bf Case of $i=k$ and $j=l$}

For brevity of explanation, we set $i=1$ and $j=2$.
Since we only consider the system at $\bsx=\bsx^*$ in the following, we drop the symbol $\bsx^*$ for brevity.
We compute $\frac{1}{\tau} \langle \hcalJ_{12} \hcalI_{ij, \bsx^*}\rangle_{\bsx^*}$ by using the method of generating functions.
Based on the transition matrix $T$, we construct a transition matrix with counting fields $T'(n_1,n_2,m)$ as
\eq{
T'_{ij}=\bcases{
T_{ij}& j\neq 1,2, \\
T_{ij}e^{n_1}&j=1 \ {\rm and} \ i\neq 2, \\
T_{ij}e^{n_1+m}&j=1 \ {\rm and} \ i=2, \\
T_{ij}e^{n_2}&j=2 \ {\rm and} \ i\neq 1, \\
T_{ij}e^{n_2-m}&j=2 \ {\rm and} \ i=1. \\
}
}
This matrix can be expressed as
\eq{
T'=\mx{
T_{11}e^{n_1}&T_{12}e^{n_2-m}&T_{13}&\cdots &T_{1n}\\
T_{21}e^{n_1+m}&T_{22}e^{n_2}&T_{23}&\cdots &T_{2n}\\
T_{31}e^{n_1}&T_{32}e^{n_2}&T_{33}&\cdots &T_{3n}\\
\vdots&\vdots&\vdots&\ddots&\vdots \\
T_{n1}e^{n_1}&T_{n2}e^{n_2}&T_{n3}&\cdots &T_{nn}\\
}.
}
Its largest eigenvalue $\lmd'$ has the information of $\hcalI$ and $\hcalJ$ in the form of
\balign{
\ft{\frac{\del}{\del m}\ln \lmd'}{n_1,n_2,m=0}&=\langle \hcalJ_{ij}\rangle, \\
\( T_{21}\ft{\frac{\del^2}{\del m \del n_1}\ln \lmd'}{n_1,n_2,m=0}-T_{12}\ft{\frac{\del^2}{\del m \del n_2}\ln \lmd'}{n_1,n_2,m=0}\)&=\langle \hcalJ_{ij}\hcalI_{ij}\rangle.
}
Since all the derivatives in this section are taken at $n_1,n_2,m=0$, in the following we drop the symbol $\ft{}{n_1,n_2,m=0}$ for brevity.

Let $\lmd^k$, $\bsp^k$, and $\bsu^k$ be the $k$-th eigenvalue and corresponding right and left eigenvectors of $T$ normalized as $\bsu^k\cdot \bsp^k=1$.
We know that $\lmd^1=1$, $\bsp^1=\bsp^{\rm ss}$, and $\bsu^1=\mx{1&1&\cdots &1}$.
Since we finally set $n_1$, $n_2$, and $m$ to zero, we regard $T'$ as a perturbed matrix from $T$ in the following form:
\eq{
T'=T+V, \hspace{15pt} V:=\mx{
T_{11}(e^{n_1}-1)&T_{12}(e^{n_2-m}-1)&0&\cdots &0\\
T_{21}(e^{n_1+m}-1)&T_{22}(e^{n_2}-1)&0&\cdots &0\\
T_{31}(e^{n_1}-1)&T_{32}(e^{n_2}-1)&0&\cdots &0\\
\vdots&\vdots&\vdots&\ddots&\vdots \\
T_{n1}(e^{n_1}-1)&T_{n2}(e^{n_2}-1)&0&\cdots &0\\
}.
}
On the basis of this observation, we expand $\lmd'$ and corresponding eigenvector $\bsp'$ from $\lmd^1=1$ and $\bsp^1=\bsp^{\rm ss}$ in terms of $n_1$, $n_2$, and $m$.
Following the standard perturbation method, we have
\eq{
\lmd'=\lmd^1+\bsu^1V\bsp^1-\sum_{k(\neq 1)}\frac{(\bsu^kV\bsp^0)(\bsu^0V\bsp^k)}{\lmd^k-1}+O(\{ n_1,n_2,m\}^3),
}
where the last term $O(\{ n_1,n_2,m\}^3)$ means that this quantity is higher or equal to the third order of $n_1$, $n_2$, and $m$.
A direct calculation yields
\balign{
\bsu^1V\bsp^1&=\[ n_1+\frac{n_1^2}{2}+T_{21}\( m+n_1m+\frac{m^2}{2}\) \] \pss_1+\[ n_2+\frac{n_2^2}{2}+T_{12}\( m+n_2m+\frac{m^2}{2}\) \] \pss_2+O(\{ n_1,n_2,m\}^3), \\
\bsu^kV\bsp^0&=\lmd^k u_1^k \pss_1n_1+\lmd^k u_2^k \pss_2n_2+(T_{21}u^k_2\pss_1-T_{12}u^k_1\pss_2)m+O(\{ n_1,n_2,m\}^2), \\
\bsu^0V\bsp^k&=p^k_1n_1+p^k_2n_2+(T_{21}p^k_1-T_{12}p^k_2)m+O(\{ n_1,n_2,m\}^2).
}
With noting the relation
\eq{
\frac{\del \lmd'}{\del m}=\lmd'\frac{\del}{\del m}\ln \lmd'=\lmd' \langle \hcalJ_{ij}\rangle=0
}
at the stalling condition $\bsx=\bsx^*$, we observe
\eq{
\frac{\del^2}{\del m\del n_1}\ln \lmd'=\frac{1}{\lmd'^2}\( \frac{\del^2 \lmd'}{\del m\del n_1}-\frac{\del \lmd'}{\del m}\frac{\del \lmd'}{\del n_1} \) =\frac{1}{\lmd'^2} \frac{\del^2 \lmd'}{\del m\del n_1}.
}
Thus, to prove \eref{IJ-zero} it suffices to show
\eqa{
T_{21}\frac{\del^2 \lmd'}{\del m\del n_1}-T_{12}\frac{\del^2 \lmd'}{\del m\del n_2}=0.
}{CJzero-mid-key}

These two derivatives are computed as
\balign{
\frac{\del^2 \lmd'}{\del m\del n_1}&=T_{21}\pss_1-\sum_{k(\neq 1)} \frac{1}{\lmd^k-1}\[ \lmd^ku^k_1\pss_1 (T_{21}p^k_1-T_{12}p^k_2)+p^k_1(T_{21}u^k_2\pss_1-T_{12}u^k_1\pss_2)\] , \\
\frac{\del^2 \lmd'}{\del m\del n_2}&=-T_{12}\pss_2-\sum_{k(\neq 1)} \frac{1}{\lmd^k-1}\[ \lmd^ku^k_2\pss_2 (T_{21}p^k_1-T_{12}p^k_2)+p^k_2(T_{21}u^k_2\pss_1-T_{12}u^k_1\pss_2)\]  .
}
Using the stalling condition $T_{12}\pss_2=T_{21}\pss_1$, the left-hand side of \eref{CJzero-mid-key} is calculated as
\balign{
T_{21}\frac{\del^2 \lmd'}{\del m\del n_1}-T_{12}\frac{\del^2 \lmd'}{\del m\del n_2}
=&T^2_{21}\pss_1-\sum_{k(\neq 1)} \frac{1}{\lmd^k-1}\[ \lmd^ku^k_1T_{21}\pss_1 (T_{21}p^k_1-T_{12}p^k_2)+T_{21}p^k_1(u^k_2-u^k_1)T_{21}\pss_1\] \nt \\
&+T_{12}T_{21}\pss_1+\sum_{k(\neq 1)} \frac{1}{\lmd^k-1}\[ \lmd^ku^k_2T_{21}\pss_1 (T_{21}p^k_1-T_{12}p^k_2)+T_{12}p^k_2(u^k_2-u^k_1)T_{21}\pss_1\] \nt \\
=&T_{21}\pss_1\[ T_{21}+T_{12}-\sum_{k\neq 1)} (u^k_1-u^k_2) (T_{21}p^k_1-T_{12}p^k_2)\] .\lb{CJzero-mid}
}
Finally, we recall the relation
\eq{
\sum_k \bsp^k\cdot \bsu^k=I,
}
which suggests
\eqa{
\sum_{k(\neq 1)}u_i^kp^k_j=\delta_{ij}-u^1_ip^1_j=\delta_{ij}-\pss_j.
}{CJzero-delta}
Plugging this relation into \eref{CJzero-mid}, we arrive at the desired result \eqref{CJzero-mid-key}:
\eq{
T_{21}+T_{12}-\sum_{k\neq 1)} (u^k_1-u^k_2) (T_{21}p^k_1-T_{12}p^k_2)=T_{21}+T_{12}-\[ T_{21}(1-\pss_1)+T_{21}\pss_1+T_{12}(1-\pss_2)+T_{12}\pss_2\] =0.
}

\bigskip

\underline{\bf Case of $i=k$ and $j\neq l$}

For brevity of explanation, we set $i=1$, $j=2$, and $l=3$.
The transition matrix with counting fields $T'(n_1,n_3,m)$ is set as
\eq{
T'=\mx{
T_{11}e^{n_1}&T_{12}e^{-m}&T_{13}e^{n_3}&\cdots &T_{1n}\\
T_{21}e^{n_1+m}&T_{22}&T_{23}e^{n_3}&\cdots &T_{2n}\\
T_{31}e^{n_1}&T_{32}&T_{33}e^{n_3}&\cdots &T_{3n}\\
\vdots&\vdots&\vdots&\ddots&\vdots \\
T_{n1}e^{n_1}&T_{n2}&T_{n3}e^{n_3}&\cdots &T_{nn}\\
}.
}
A direct calculation yields
\balign{
\bsu^1V\bsp^1&=\[ n_1+\frac{n_1^2}{2}+T_{21}\( m+n_1m+\frac{m^2}{2}\) \] \pss_1+\[ T_{12}\( m+\frac{m^2}{2}\) \] \pss_2+\[ n_3+\frac{n_3^2}{2}\] \pss_3 +O(\{ n_1,n_3,m\}^3), \\
\bsu^kV\bsp^0&=\lmd^k u_1^k \pss_1n_1+\lmd^k u_3^k \pss_3n_3+(T_{21}u^k_2\pss_1-T_{12}u^k_1\pss_2)m+O(\{ n_1,n_2,m\}^2), \\
\bsu^0V\bsp^k&=p^k_1n_1+p^k_3n_3+(T_{21}p^k_1-T_{12}p^k_2)m+O(\{ n_1,n_2,m\}^2).
}
Again, the cross correlation term $\langle \hcalJ_{12}\hcalI_{13, \bsx^*}\rangle_{\bsx^*}$ is written as
\eq{
\langle \hcalJ_{21}\hcalI_{31, \bsx^*}\rangle_{\bsx^*}=T_{31}\frac{\del^2 \lmd'}{\del m\del n_1}-T_{13}\frac{\del^2 \lmd'}{\del m\del n_3}
}
Two derivatives are computed as
\balign{
\frac{\del^2 \lmd'}{\del m\del n_1}&=T_{21}\pss_1-\sum_{k(\neq 1)} \frac{1}{\lmd^k-1}\[ \lmd^ku^k_1\pss_1 (T_{21}p^k_1-T_{12}p^k_2)+p^k_1(T_{21}u^k_2\pss_1-T_{12}u^k_1\pss_2)\] , \\
\frac{\del^2 \lmd'}{\del m\del n_3}&=-\sum_{k(\neq 1)} \frac{1}{\lmd^k-1}\[ \lmd^ku^k_3\pss_3 (T_{21}p^k_1-T_{12}p^k_2)+p^k_3(T_{21}u^k_2\pss_1-T_{12}u^k_1\pss_2)\] .
}
Another cross term $\langle \hcalJ_{13}\hcalI_{12, \bsx^*}\rangle_{\bsx^*}$ is expressed in a similar manner.
Using the stalling conditions $T_{12}\pss_2=T_{21}\pss_1$ and $T_{13}\pss_3=T_{31}\pss_1$, we have
\balign{
\langle \hcalJ_{21}\hcalI_{31, \bsx^*}\rangle_{\bsx^*}+\langle \hcalJ_{31}\hcalI_{21, \bsx^*}\rangle_{\bsx^*}=
&T_{31}T_{21}\pss_1-\sum_{k(\neq 1)} \frac{T_{31}}{\lmd^k-1}\[ \lmd^ku^k_1\pss_1 (T_{21}p^k_1-T_{12}p^k_2)+p^k_1(T_{21}u^k_2\pss_1-T_{12}u^k_1\pss_2)\] \nt \\
&+\sum_{k(\neq 1)} \frac{T_{13}}{\lmd^k-1}\[ \lmd^ku^k_3\pss_3 (T_{21}p^k_1-T_{12}p^k_2)+p^k_3(T_{21}u^k_2\pss_1-T_{12}u^k_1\pss_2)\]  \nt \\
&+T_{31}T_{21}\pss_1-\sum_{k(\neq 1)} \frac{T_{21}}{\lmd^k-1}\[ \lmd^ku^k_1\pss_1 (T_{31}p^k_1-T_{13}p^k_3)+p^k_1(T_{31}u^k_3\pss_1-T_{13}u^k_1\pss_3)\] \nt \\
&+\sum_{k(\neq 1)} \frac{T_{12}}{\lmd^k-1}\[ \lmd^ku^k_2\pss_2 (T_{31}p^k_1-T_{13}p^k_3)+p^k_2(T_{31}u^k_3\pss_1-T_{13}u^k_1\pss_3)\] \nt \\
=&2T_{31}T_{21}\pss_1-\sum_{k(\neq 1)}[(u_1^k-u_3^k)(T_{21}p_1^k-T_{12}p_2^k)T_{31} + (u_1^k-u_2^k)(T_{31}p_1^k-T_{13}p_3^k)T_{21} ] \pss_1.
}
Using \eref{CJzero-delta}, the right-hand side turns out to be zero.

\bigskip

\underline{\bf Case of $i\neq k$ and $j\neq l$}

For brevity of explanation, we set $i=1$, $j=2$, $k=3$, and $l=4$.

Calculation for this case is essentially the same as the case of $i=k$ and $j\neq l$.
With the same notation, the result is expressed as
\balign{
\langle \hcalJ_{21}\hcalI_{43, \bsx^*}\rangle_{\bsx^*}+\langle \hcalJ_{43}\hcalI_{21, \bsx^*}\rangle_{\bsx^*}=
&-\sum_{k(\neq 1)} \frac{T_{43}}{\lmd^k-1}\[ \lmd^ku^k_3\pss_3 (T_{21}p^k_1-T_{12}p^k_2)+p^k_3(T_{21}u^k_2\pss_1-T_{12}u^k_1\pss_2)\] \nt \\
&+\sum_{k(\neq 1)} \frac{T_{34}}{\lmd^k-1}\[ \lmd^ku^k_4\pss_4 (T_{21}p^k_1-T_{12}p^k_2)+p^k_4(T_{21}u^k_2\pss_1-T_{12}u^k_1\pss_2)\]  \nt \\
&-\sum_{k(\neq 1)} \frac{T_{21}}{\lmd^k-1}\[ \lmd^ku^k_1\pss_1 (T_{43}p^k_3-T_{34}p^k_4)+p^k_1(T_{43}u^k_4\pss_3-T_{34}u^k_3\pss_4)\] \nt \\
&+\sum_{k(\neq 1)} \frac{T_{12}}{\lmd^k-1}\[ \lmd^ku^k_2\pss_2 (T_{43}p^k_3-T_{34}p^k_4)+p^k_2(T_{43}u^k_4\pss_3-T_{34}u^k_3\pss_4)\] \nt \\
=&-\sum_{k(\neq 1)}(u_3^k-u_4^k)(T_{21}p_1^k-T_{12}p_2^k)T_{43}\pss_3 + (u_1^k-u_2^k)(T_{43}p_3^k-T_{34}p_4^k)T_{21} \pss_1.
}
Using \eref{CJzero-delta}, the right-hand side turns out to be zero.

\bigskip\noindent
{\bf \large B. Derivations of two fluctuation theorems on partial entropy productions}
\midskip

\smallskip

\underline{\bf Proof of \eref{SS-FT}}

We shall prove the fluctuation theorem
\eq{
\la e^{-\hsgm_{\Omega}}\ra_{\bsx}^\tau=1,
}
with the partial entropy production for a set of edges $\Omega$
\balign{
\hsgm_{\Omega}:=&\sum_{e\in \Omega}a_e(\bsx)\hcalJ_e-J_e(\bsx)\hcalC_{e,\bsx}.
}
Here, $a_{ij}$ the total force associated with the edge
\eq{
a_{ij}(\bsx):=\ln \frac{R_{ij}(\bsx)\pss_{j}(\bsx)}{R_{ji}(\bsx)\pss_{i}(\bsx)}.
}
Since the parameter $\bsx$ is fixed in this discussion, we drop $\bsx$ dependence in the following.

To prove this, we introduce another transition rate $P'$ defined as
\eq{
R'_{ij}=
\begin{cases}
R_{ij} &: (i,j) \in \Omega, \\
& \\
\dis\frac{R_{ji}\pss_i}{\pss_j} &: (i,j) \notin \Omega.
\lb{way-reverse}
\end{cases}
}
Its escape rate $e'_{j}:=-R'_{jj}=\sum_{i(\neq j)} R'_{ij}$ is calculated as
\balign{
e '_{j}
=& \sum_{i:(i,j)\notin \Omega}\frac{R_{ji}\pss_i}{\pss_j} + \sum_{i:(i,j)\in \Omega}R_{ij} 
= \sum_{i:(i,j)\notin \Omega}R_{ij}+\frac{R_{ji}\pss_i-R_{ij}\pss_j}{\pss_j} + \sum_{i:(i,j)\in \Omega} R_{ij}
=e _{j}+\frac{J_{j,\Omega^\mr{c}}}{\pss_j},
}
where $j$ runs all states neighboring $j$, and we defined $J_{j,\Omega^\mr{c}}:=\sum_{i:(i,j)\notin \Omega}J_{ji}= \sum_{i:(i,j)\notin \Omega} R_{ji}\pss_i-R_{ij}\pss_j$ as the current to the state $j$ only through edges in $\Omega^{\rm c}$.
We define $J_{j,\Omega}$ in a similar manner.

We next compute the contribution to $e^{-\hsgm_\Omega}$ from each single jump $i\to j$.
For the case of $(i,j)\in \Omega$, we have
\eq{
R_{ji}e^{-a_{ji}\hcalJ_{ji}}
=R_{ji}\cdot \frac{R_{ij}}{R_{ji}}\frac{\pss_j}{\pss_i} 
=R'_{ij}\frac{\pss_j}{\pss_i}, 
}
and for the case of $(i,j) \notin \Omega$, we again have
\eq{
R_{ji}e^{-a_{ji}\hcalJ_{ji}}
=R_{ji}\cdot 1 
=R'_{ij}\frac{\pss_j}{\pss_i}.
}
Hence,
\eq{
R_{ji}e^{-a_{ji}\hcalJ_{ji}}=R'_{ij}\frac{\pss_j}{\pss_i}
}
holds for any transition $i\to j$.
Using this, we arrive at the desired relation
\balign{
\la e^{-\hsgm_\Omega}\ra _{\bsx}^\tau
=&\int d\Gamma \pss_{w^{0}}\cdot\( \prod_{i=1}^{N}R_{w^i,w^{i-1}} \) \( \prod_{i=0}^Ne^{-\int_{t^{i}}^{t^{i+1}}e _{w^{i}}dt}\) \cdot e^{-\sum_{e\in \Omega}a_e(\bsx)\hcalJ_e+J_e(\bsx)\hcalC_{e,\bsx} } \nt \\
=&\int d\Gamma \pss_{w^{0}} \cdot\( \prod_{i=1}^{N} \frac{\pss_{w^i}}{\pss_{w^{i-1}}}R'_{w^{i-1}w^{i}}\) \( \prod_{i=0}^Ne^{-\int_{t^{i}}^{t^{i+1}}e _{w^{i}} dt-\int_{t^i}^{t^{i+1}}J_{w^i, \Omega^\mr{c}}/\pss_{w^i}dt}\) \nt \\
=&\int d\Gamma \pss_{w^{N}} \cdot\( \prod_{i=1}^{N}R'_{w^{i-1}w^{i}} \) \( \prod_{i=0}^Ne^{-\int_{t^{i}}^{t^{i+1}}e ' _{w^{i}} dt}\) \nt \\
=&\int d\Gamma P'(\Gamma^\dagger) \nt \\
=&1.
}
Here, $P'(\Gamma)$ is the path probability with the transition rate $R'$, and $\Gamma$ is a trajectory where the number of jumps is $N$ and the $i$-th jump occurs from a state $w^{i-1}$ to another state $w^i$ at time $t^i$.
In the third line, we used $J_{w, \Omega}+J_{w, \Omega^\mr{c}}=\frac{d}{dt}\pss_w=0$.

\bigskip

\underline{\bf Proof of \eref{APE-FT}}

We next prove the fluctuation theorem
\eq{
\langle e^{-\hsgm_\Omega^{\rm I}}\rangle_{\bsx|\bsp(\bsx^*)}^\tau =1
}
with the informed partial entropy production 
\eq{
\hsgm_\Omega^{\rm I}:=\sum_{e\in \Omega}\hcalJ_e \tlF^{\bsx}_{e}.
}
Here, $\tlF^{\bsx}_{e}$ is an {\it informed force} defined as
\eqa{
\tlF^{\bsx}_{ij}:=\ln \frac{R_{ij}(\bsx)\pss_{j}(\bsx^*)}{R_{ji}(\bsx)\pss_{i}(\bsx^*)}.
}{def-informed-force}
In the following, we abbreviate $\pss_{i}(\bsx^*)$ as $p^*_i$ for brevity.

We introduce another transition rate ${R'}({\bsx})$ such that
\eq{
{R'}_{ij}({\bsx})=
\bcases{
R_{ij}({\bsx}) & (i,j)\in \Omega, \\
\frac{R_{ji}({\bsx})p_{i}^*}{p_j^*} & (i,j)\notin \Omega .
}
}
This transition rate is the same as the original transition rate if $(i,j)\in \Omega$, and is the dual transition rate at $\bsx=\bsx^*$ if $(i,j)\notin \Omega$ (we note $R_{ji}(\bsx)=R_{ji}(\bsx^*)$ for $(i,j)\notin \Omega$).
Notably, the escape rate of ${R'}({\bsx})$ denoted by ${e'}^{\bsx}$ is the same as that of $R({\bsx})$, which can be shown as
\eq{
{e'}_j^{\bsx}=\sum_{i(\neq j)}{R'}_{ij}({\bsx}) 
=\sum_{i: (i,j)\in \Omega} R_{ij}({\bsx}) + \sum_{i: (i,j)\notin \Omega} \frac{R_{ji}(\bsx)p_{i}^*}{p_j^*} 
=\sum_{i: (i,j)\in \Omega} R_{ij}({\bsx}) + \sum_{i: (i,j)\notin \Omega}R_{ij}(\bsx)
=e_j^{\bsx},
}
where the second equality follows from
\eq{
 \sum_{i: (i,j)\notin \Omega} \frac{R_{ji}(\bsx)p_{i}^*}{p_j^*}
= \sum_{i: (i,j)\notin \Omega} \frac{R_{ji}(\bsx^*)p_{i}^*}{p_j^*}
= \sum_{i(\neq j)}{R}_{ij}({\bsx^*})-\sum_{i: (i,j)\in \Omega} \frac{R_{ji}(\bsx^*)p_{i}^*}{p_j^*} 
= \sum_{i: (i,j)\notin \Omega}{R}_{ij}({\bsx^*}) 
= \sum_{i: (i,j)\notin \Omega}R_{ij}(\bsx).
}
In the second equality, we used the stationary condition $\sum_{i(\neq j)}R_{ij}(\bsx^*)p_j^*-R_{ji}(\bsx^*)p_i^*$, and in third equality we used the stalling condition $R_{ji}(\bsx^*)p_{i}^*=R_{ij}({\bsx^*})p_j^*$ for $(i,j)\in \Omega$.

We now consider the contribution to $e^{-\hsgm_\Omega^{\rm I}}$ from each single jump $w'\to w$.
If $(i,j)\in \Omega$, we have
\eq{
R_{ij}(\bsx)e^{-\tlF^{\bsx}_{ij}\hcalJ_{ij}}={R'}_{ji}(\bsx)\frac{p^*_{i}}{p^*_j},
}
and if $(i,j)\notin \Omega$, we again have
\eq{
R_{ij}(\bsx)={R'}_{ji}(\bsx)\frac{p^*_{i}}{p^*_j}.
}
We then obtain 
\balign{
P^{\bsx}(\Gamma) e^{-\hsgm^{\rm I}_{\Omega}}
=&p^*_{w^0}\cdot \(\prod_{i=1}^N \frac{p^*_{w^i}}{p^*_{w^{i-1}}}\) \cdot \( \prod_{i=1}^N {R'}_{w^{i-1}, w^i}({\bsx})\) \cdot \( \prod_{i=0}^Ne^{-(t^{i+1}-t^i)e'_{w^i}}\)  \nt \\
=&p^*_{w^N}\cdot \( \prod_{i=1}^N {R'}_{w^{i-1}, w^i}({\bsx})\) \cdot \( \prod_{i=0}^Ne^{-(t^{i+1}-t^i)e'_{w^i}}\)  \nt \\
=&{P'}^{\bsx}(\Gamma^\dagger),
}
where the initial distribution of ${P'}^{\bsx}$ is set to $\bsp^*$.
By integrating both sides with $\Gamma$, we obtain \eref{APE-FT}.

\bigskip\noindent
{\bf \large C. Derivation of fluctuation-response relation from integral fluctuation theorem}
\midskip

We here briefly explain a standard technique to derive a fluctuation-response relation from an integral fluctuation theorem.
Suppose that the entropy production is written as
\eq{
\hsgm=F\hcalJ^F.
}
With regarding $F$ as a small parameter, we perform the Taylor expansion of the exponential function in the integral fluctuation theorem as
\eq{
1=\langle e^{-\hsgm}\rangle_F=1-\langle F\hcalJ^F\rangle_F+\frac12 \langle (F \hcalJ^F)^2\rangle_F+\cdots.
}
Since $\langle \cdot \rangle_F$ is also a function of $F$, this bracket can also be expanded as
\eq{
\langle \cdot \rangle_F=\langle \cdot \rangle_0+F \ft{\frac{\del}{\del F}\langle \cdot \rangle_F}{F=0}+\ft{\frac{F^2}{2}\frac{\del^2}{\del F^2}\langle \cdot \rangle_F}{F=0}\cdots.
}
These two expansions suggest
\balign{
1=&1-F \langle \hcalJ^F\rangle_0+F^2 \ft{\frac{\del}{\del F}\langle \hcalJ^F \rangle_F}{F=0} +\frac{F^2}{2} \langle (\hcalJ^F)^2\rangle_0+O(F^3).
}
Comparing the coefficients of $F^2$, we find the fluctuation-response relation:
\eq{
 \ft{\frac{\del}{\del F}\langle \hcalJ^F \rangle_F}{F=0}=\frac{1}{2} \langle (\hcalJ^F)^2\rangle_0.
}

\end{widetext}
\end{document}